\documentclass[journal=aamick,manuscript=article]{achemso}

\usepackage[dvipsnames,table]{xcolor}
\usepackage[version=3]{mhchem} %
\usepackage{miller}
\usepackage{subcaption}
\usepackage{multirow}
\usepackage{color}
\usepackage{textgreek}

\usepackage{amsmath,amssymb}
\graphicspath{{figs/}}

\usepackage{pdfpages}

 \usepackage[font=small]{caption}

\usepackage[%
  breaklinks,             %
  bookmarks = false, %
  pdfpagemode   = UseNone,%
  pdfstartview  = FitH,     %
  pdfstartpage  = 1,      %
  colorlinks    = true,   %
]{hyperref}

\newcommand\myshade{85}
\colorlet{mylinkcolor}{YellowOrange}
\colorlet{myurlcolor}{Aquamarine}
\colorlet{mycitecolor}{violet}

\hypersetup{
  linkcolor  = mylinkcolor!\myshade!black,
  citecolor  = mycitecolor!\myshade!black,
  urlcolor   = myurlcolor!\myshade!black,
  colorlinks = true,
}



\author{Moin Uddin Maruf}
\affiliation{Department of Mechanical Engineering, Texas Tech University, Lubbock, Texas 79409, USA}
\author{Sungmin Kim}
\affiliation{Samsung Advanced Institute of Technology, Samsung Electronics, Suwon 16678, Republic of Korea}
\author{Zeeshan Ahmad}
\affiliation{Department of Mechanical Engineering, Texas Tech University, Lubbock, Texas 79409, USA}
\email{zeeahmad@ttu.edu}

\title{Equivariant Machine Learning Interatomic Potentials with Global Charge Redistribution}


\begin{document}

\begin{center}
\end{center}

\begin{abstract}

Machine learning interatomic potentials (MLIPs) provide a computationally efficient alternative to quantum mechanical simulations for predicting material properties. Message-passing graph neural networks, commonly used in these MLIPs, rely on local descriptor-based symmetry functions to model atomic interactions. However, such local descriptor-based approaches struggle with systems exhibiting long-range interactions, charge transfer, and compositional heterogeneity. In this work, we develop a new equivariant MLIP incorporating long-range Coulomb interactions  through explicit treatment of electronic degrees of freedom, specifically global charge distribution within the system. This is achieved using a charge equilibration scheme based on predicted atomic electronegativities. We systematically evaluate our model across a range of benchmark periodic and non-periodic datasets, demonstrating that it outperforms both short-range equivariant and long-range invariant MLIPs in energy and force predictions. Our approach enables more accurate and efficient simulations of systems with long-range interactions and charge heterogeneity, expanding the applicability of MLIPs in computational materials science.

\end{abstract}

Machine learning has revolutionized computational materials science by offering highly accurate and predictive surrogate models that significantly accelerate quantum mechanical calculations~\cite{Ramprasad2017,Ong2019}. Early applications focused on building machine learning models to predict properties of functional materials such as mechanical properties and ion conductivity from the structure directly~\cite{pilania2013accelerating,liuMaterialsDiscoveryDesign2017,wangMachineLearningMaterials2020}. These models  facilitated the design and discovery of novel materials for applications such as energy storage and conversion, healthcare, and computing~\cite{ahmadMachineLearningEnabled2018,venturiMachineLearningEnabled2020,freyMachineLearningEnabledDesign2020,Ekins2019}.
Recently, machine learning interatomic potentials (MLIPs) which operate at a more fundamental level, serving as direct surrogates for quantum mechanical simulations such as density functional theory (DFT) have been gaining traction for more rigorous and accurate prediction of material properties~\cite{Chen2024,Qi2021Bridging,Deringer2019,behlerPerspectiveMachineLearning2016}. When trained on large datasets from simulations, they can predict the potential energy surface of atomic structures with remarkable accuracy. These potentials enable large-scale simulations of materials at a reduced computational cost. This efficiency allows exploration of larger systems and longer time scales, making it feasible to investigate phenomena that were previously computationally prohibitive.

Current state-of-the-art MLIPs employ message parsing graph neural networks architectures, operating under the assumption that atomic properties depend on their local chemical environment, neglecting the influence of atoms located beyond a cutoff radius. This assumption implies that the total energy of a system is the sum of all local atomic contributions, which allows MLIPs to scale linearly with system size. Since these cutoff-based short-range MLIPs neglect long-range interactions such as Coulomb and dispersion, they may be very successful in describing properties of bulk systems~\cite{homogeneousBulkSystem2021}, but they fail for many others systems and applications such as liquid-vapor interfaces, dielectric response, dilute ionic solutions and interactions between gas phase molecules \cite{InteractionBetweenGasPhaseMolecules2023,dielectricResponse1-2008, dielectricResponse2-2008}.

Long-range MLIPs have recently emerged as a promising approach to overcome the limitations imposed by the finite cutoff radius in short-range MLIPs~\cite{anstineMachineLearningInteratomic2023,koGeneralPurposeMachineLearning2021}. Long-range electrostatic correction for ionic systems based on fixed charges have been proposed \cite{GaussianApproxPotential2010,fixedCharge2019}. This approach has been improved  by constructing environment-dependent atomic charges using a set of atomic neural networks \cite{BehlerHDNN2011,PESwaterDimer2012}. However, since the atomic charges are predicted as a function of the local descriptors, capturing the correct global charge distribution is not always guaranteed.

To allow a global charge redistribution over the system, the charge equilibration via neural network technique (CENT) has been proposed in which the charge-dependent total energy expression is minimized~\cite{ghasemi2015interatomic2015}. This approach conserves the total charge and uses a set of neural networks to predict electronegativity as a function of local descriptors. Then, a series of coupled linear equations are solved for atomic charges, allowing charge redistribution throughout the system. Though CENT scheme has been successfully applied to implement MLIPs for ionic systems \cite{ghasemi2015interatomic2015, faraji2017high, hafizi2017neural}, it may not work well for systems with the coexistance of covalent and ionic interactions as the functional form of total energy lacks the description for covalent bonding. ~\citet{koFourthgenerationHighdimensionalNeural2021a} developed fourth generation high-dimensional neural network potentials (4G-HDNNP) by combining local-descriptor based methods and CENT scheme. They use two sets of invariant neural networks: one for predicting atomic electronegativities and the other for short-range atomic energies. Atomic charges are predicted via charge equilibration scheme and compared with Hirshfeld charges to update the electronegativities. The predicted charges along with atomic descriptor functions are further used to predict energy and forces.
Overall, MLIPs that incorporate electronic degrees of freedom, such as charges, spin, and magnetism, in addition to the potential energy surface, have shown significant improvements in predictive accuracy compared to those without this information~\cite{unkeSpookyNetLearningForce2021, dengCHGNetPretrainedUniversal2023}.

Equivariant MLIPs such as neural equivariant interatomic potential (NequIP)~\cite{batznerEquivariantGraphNeural2022} and MACE~\cite{batatiaMACEHigherOrder2022} represent the next frontier in advancing the capabilities of MLIPs, offering significantly improved data efficiency. By leveraging symmetries such as rotational, translational, and permutational equivariance, these MLIPs offer enhanced expressivity and can learn effectively from limited data while maintaining high accuracy~\cite{thomasTensorFieldNetworks2018}. These potentials have rapidly become the state-of-the-art for predicting material properties, outperforming traditional potentials in accuracy and efficiency~\cite{thomasTensorFieldNetworks2018,batznerEquivariantGraphNeural2022,batatiaMACEHigherOrder2022,musaelianLearningLocalEquivariant2023}. 
Recently, long-range interactions have been incorporated into equivariant MLIPs, demonstrating improved accuracy on benchmark systems such as \ce{Au2} cluster on MgO substrate\cite{chengLatentEwaldSummation2024,kimLearningChargesLongrange2024,gaoEnhancingUniversalMachine2024}. However, the utility of atomic charges which are not observables and the validity of global charge equilibration  remains a subject of debate.

In this work, we develop and implement a long-range equivariant MLIP based on NequIP and systematically compare the performance of  short-range equivariant, long-range invariant, and long-range equivariant MLIPs for a set of benchmark systems. We demonstrate that our long-range equivariant MLIP outperforms the others in prediction of energies and forces for a diverse set of systems.
The MLIP uses a charge equilibration scheme to predict the global charge redistribution among the atoms. The charge distribution combined with local features allows accurate predictions of short and long-range energies and forces.
Further, our approach enhances the capability of equivariant models by allowing simultaneous training and property predictions for systems in different charge states. The atomic charges can either be explicitly trained using reference DFT charges when included in the loss function or treated as internal parameters that function as response charges~\cite{kimLearningChargesLongrange2024}.
Our model paves the way for more accurate and versatile simulations of complex materials systems by combining equivariance for data efficiency with physics-based long-range interactions.

\section*{Results}
\subsection*{Equivariant MLIP architecture with global charge redistribution}

\begin{figure}[htbp]
    \includegraphics[width=\textwidth]{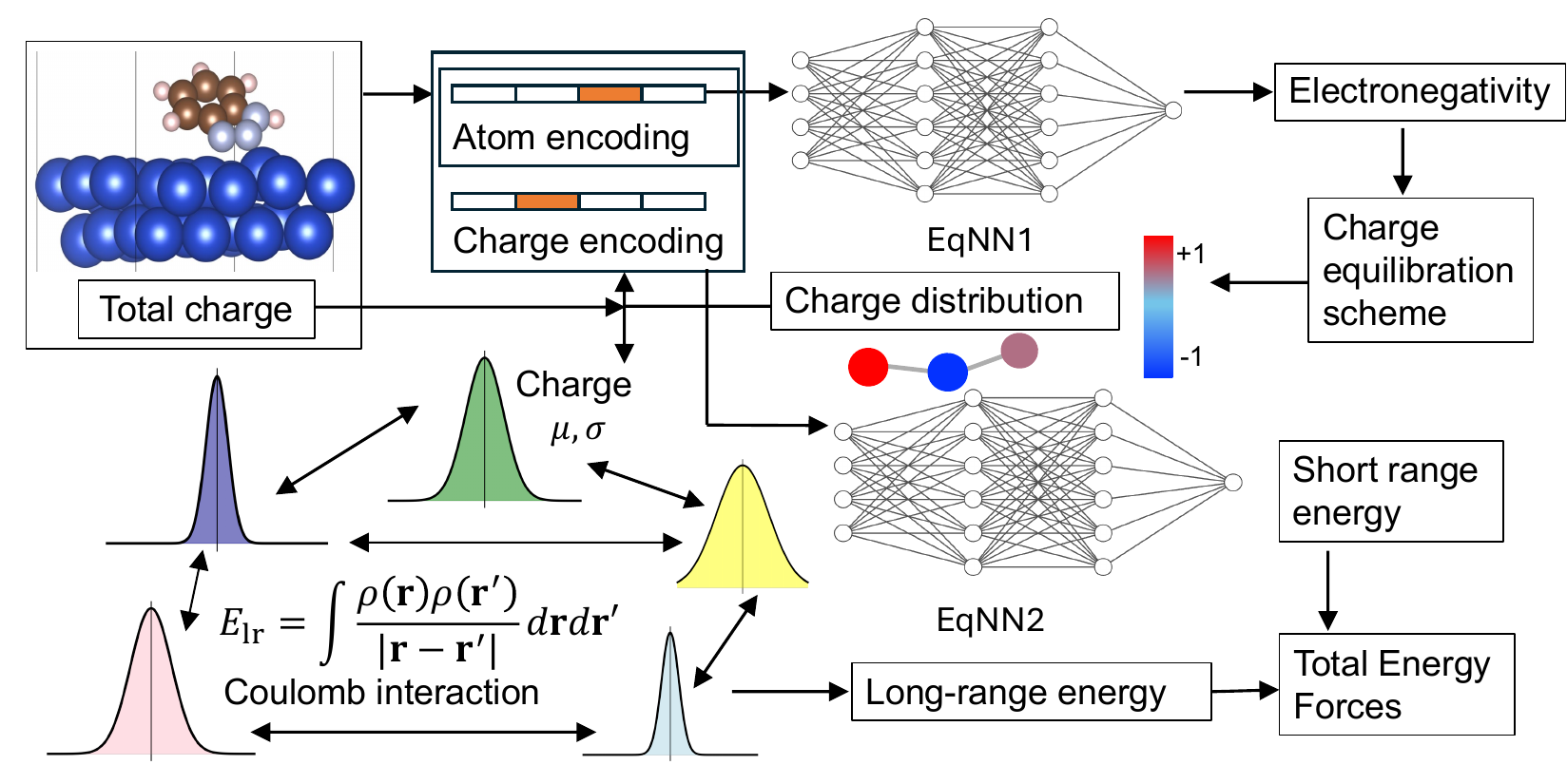}
    \caption{Architecture of the developed NequIP-LR model, comprising two equivariant neural networks, EqNN1 and EqNN2. EqNN1 predicts the electronegativity for each atom, facilitating the calculation of charge distribution. EqNN2 predicts short-range energy based on the calculated charges and atomic positions. The total energy is obtained by summing the short-range energy from EqNN2 and the long-range energy calculated using the Coulomb integral.}
    \label{fig:architecture}
\end{figure}

\autoref{fig:architecture} shows the architecture of our equivariant long-range MLIP with global charge redistribution. Our architecture combines the equivariance property of NequIP~\cite{batznerEquivariantGraphNeural2022} with the long-range treatment of 4G-HDNNP, hence, we refer to it as NequIP-LR~\cite{koFourthgenerationHighdimensionalNeural2021a}.
The MLIP predicts long-range and short-range energies separately similar to 4G-HDNNP\cite{koFourthgenerationHighdimensionalNeural2021a}. 
Our architecture consists of two equivariant neural networks, EqNN1 and EqNN2. 
The atoms are encoded as vectors through one-hot encoding according to atom type. Encoded atoms and their coordinates are passed through EqNN1 to predict the electronegativity of each atom. The atomic electronegativities are used within a charge equilibration scheme~\cite{rappeChargeEquilibrationMolecular1991a} to calculate the global charge distribution \cite{otero2014critic2} in the structure. 
The charge distribution is used to calculate the long-range interaction energy through the Coulomb integral. Ewald summation is used to sum up the integral for periodic systems.

In the next step, a new representation of the atoms that encodes both the atom type and charge is passed through EqNN2, to predict the short-range energy of the structure. The atomic charges are encoded by applying a Gaussian filter and concatenated with the atom type encoding.
The total energy is the sum of the long-range and short-range energy of the system. Forces are obtained as the derivative of the total energy through automatic differentiation. The loss function during training includes errors in energies, forces, and charges:
\begin{equation}
    \mathcal{L} = \alpha_E || \hat{E} - E_{\text{DFT}}|| + \alpha_F \frac{1}{3N_{\text{at}}} \sum_{i=1}^{N_{\text{at}}}  \sum_{j=1}^3 || \hat{F}_j - F_j || + \alpha_Q \sum_{i=1}^{N_{\text{at}}} || \hat{Q}_i - Q_i  ||
\end{equation}

where $\hat{\cdot}$ denotes predicted quantities. We use the Hirshfeld charges as reference values for learning charge distribution during training~\cite{hirshfeldBondedatomFragmentsDescribing1977a}. However, our model does not require explicit learning of atomic charges in which case the loss function consists of errors in energies and forces only. 
In these cases, atomic charges act as internal parameters for computing energies and forces, similar to the response charges in the latent Ewald summation scheme~\cite{chengLatentEwaldSummation2024} used by \citet{kimLearningChargesLongrange2024}. A key distinction in our approach is that these response charges are computed using a charge equilibration scheme based on atomic electronegativity.

To enable a meaningful comparison, we use NequIP~\cite{batznerEquivariantGraphNeural2022} as the baseline equivariant model. On top of this model, we incorporate the atomic charge distribution and long-range interactions.  The total number of convolutional layers is kept the same in both NequIP and NequIP-LR to allow a fair comparison, i.e., the sum of the number of layers in EqNN1 and EqNN2 is equal to the number of short-range layers in NequIP.
Following the recommendations for NequIP, we set the values of $\alpha_E = 1 $ and $\alpha_F=N_{\text{at}}^2$. The value of $\alpha_Q$ was varied and is discussed in the results. We also compare our results with the 4G-HDNNP which is an invariant long-range MLIP~\cite{koFourthgenerationHighdimensionalNeural2021a}.

\subsection*{Performance on benchmark datasets}
We test our model on several benchmark systems that exhibit long-range/nonlocal interactions or charge transfer. In addition,  our datasets include structures that can exist in different charge states.

\subsubsection*{\ce{Au2} cluster on MgO\hkl(001) surface}

\begin{figure}[htbp]
    \centering
    \includegraphics[width=\textwidth]{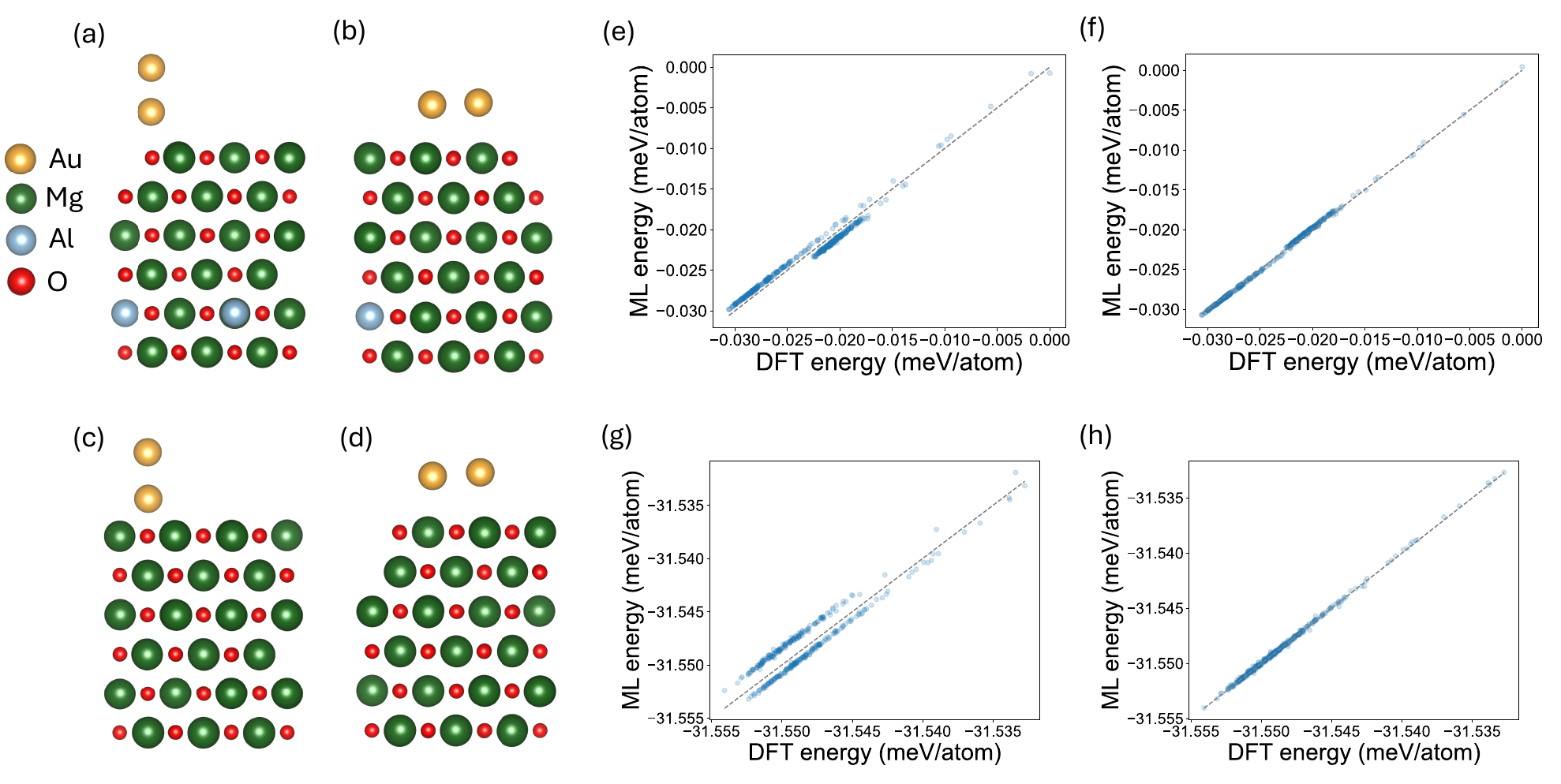}
    \caption{\ce{Au2} cluster in the non-wetting geometry on (a) Al-doped and (c) undoped, wetting geometry in (b) Al-doped and (d) undoped MgO\hkl(001) surface of (3x3) supercell. DFT optimizations of these structures show that wetting geometry is more stable in Al doped MgO\hkl(001) surface whereas non-wetting geometry is preferred on pristine MgO\hkl(001) surface. Parity plot of the energies of Al-doped structures obtained from (e) NequIP and (f) NequIP-LR and the energies of undoped structures obtained from (g) NequIP and (h) NequIP-LR.
    }
    \label{fig:Au2-MgO-Al}
\end{figure}

This dataset, generated by \citet{koFourthgenerationHighdimensionalNeural2021a}, contains \ce{Au2} cluster on undoped or Al-doped \ce{MgO}\hkl(001) surface as shown in \autoref{fig:Au2-MgO-Al}a-d. In the doped structures (\autoref{fig:Au2-MgO-Al}a and b) comprising half of the dataset, Al atoms are located within the MgO substrate at distances sufficiently far from the \ce{Au2} cluster. This dataset provides a good platform to test whether an MLIP can simulate charge transfer from Al to Au and capture interactions beyond the cutoff distance. The charge transfer from Al modifies the stable \ce{Au2} morphology on the MgO substrate by changing the stable geometry of the \ce{Au2} cluster\cite{mammen2018inducing}. All structures have zero total charge. 

The \ce{Au2} cluster on MgO\hkl(001) surface has two main adsorption geometries, an upright ``non-wetting" orientation of the dimer and another parallel to the surface in a ``wetting" configuration. DFT optimization of these structures shows that the presence of dopant atoms changes the relative stability of the structure: wetting geometry is more stable in Al-doped MgO\hkl(001) surface, whereas non-wetting geometry is more stable with undoped MgO\hkl(001) surface~\cite{mammenTuningMorphologyGold2011}. The Al dopant is introduced in the fifth layer, resulting in a distance of more than 10 {\AA} from the Au atoms.

For this dataset, 4G-HDNNP gives a force root mean squared error (RMSE) of 66 meV/{\AA}.  NequIP considerably reduces the RMSE to 34.32 meV/{\AA} while NequIP-LR further reduces the RMSE to 21.99 meV/{\AA}. For energies, 4G-HDNNP gives an RMSE of 0.219 meV/atom which increases to 1.030 meV/{\AA} using NequIP. NequIP-LR reduces it to 0.173 meV/{\AA}.

The accuracy of NequIP and NequIP-LR on the test set is compared using the parity plots in \autoref{fig:Au2-MgO-Al}e-h. The training parameters are provided in the Supporting Information. Panels (e) and (f) show the predicted energies for the Al-doped structure using NequIP and NequIP-LR, respectively, while panels (g) and (h) present the results for the undoped structure. NequIP often overestimates or underestimates the energies relative to DFT, whereas NequIP-LR provides more accurate predictions, as seen by its closer alignment with the parity line. The parity plots for forces are shown in Fig. S5.

\subsubsection*{Benzotriazole (BTA) on Cu\hkl(111) surface}
The BTA-Cu dataset contains structures of BTA molecule on Cu \hkl(111) surface in different configurations. BTA is a corrosion inhibitor used to prevent the corrosion of Cu surface and flatten the wafer surface after chemical mechanical planarization~\cite{tanEnvironmentallySustainableCorrosion2022}. The position of BTA molecule relative to the Cu surface results in a high degree of variability in the charge distribution within the system, making it challenging to capture the electrostatic interaction by a charge-agnostic MLIP. Further, the structure possesses different configurations where BTA molecule can be chemisorbed or physisorbed on the surface at different angles to the Cu surface\cite{kokalj2015ab,gattinoni2015understanding}. In addition, the Cu atoms  can have multiple possible oxidation states with different charges. 

\begin{figure}
    \centering
    \includegraphics[width=0.8\textwidth]{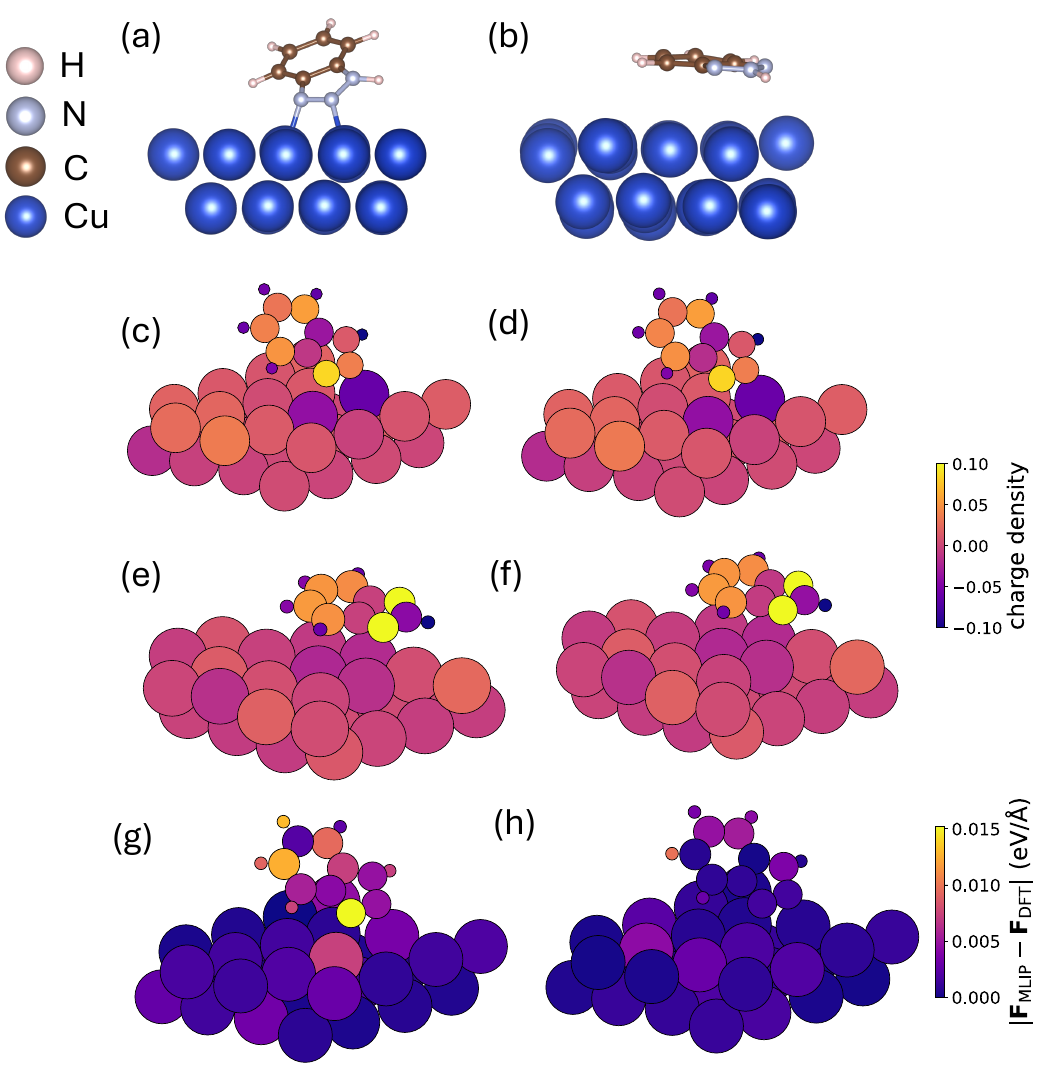}
    \caption{Benzotriazole (BTA) molecule in (a) chemisorbed and (b) physisorbed configuration on a Cu\hkl(111) surface. The varying distances between the atoms of BTA and the Cu\hkl(111) surface result in a heterogeneous charge distribution near the interface. 
    Charge distribution on the BTA molecule on  Cu\hkl(111) surface represented through 
Hirshfeld charges for chemisorbed configurations obtained using (c) DFT and (d) NequIP-LR and for physisorbed configurations obtained using (e) DFT and (f) NequIP-LR. NequIP-LR predicts the charge distribution similar to DFT. In the chemisorbed configuration, Cu atoms near the BTA molecule have acquire a more negative charge. Comparison of force errors obtained from (g) NequIP and (h) NequIP-LR. The interfacial atoms exhibit higher force errors with NequIP, which is significantly reduced by NequIP-LR. This analysis shows the impact of incorporating charge along with long-range interaction for predicting the interaction of molecules with metal surfaces.}
    \label{fig:Cu-BTA}
\end{figure}

We generated the BTA-Cu dataset using ab initio molecular dynamics (MD) simulation trajectories for chemisorbed and physisorbed configurations separately as shown in \autoref{fig:Cu-BTA}a and b respectively, each contributing to half of the dataset with a total of 400 steps. In the chemisorbed configuration, the smallest Cu-N distance is ${\sim}$ 2.1 {\AA} and the angle between Cu surface and BTA is ${\sim}$$71^\circ$  while the physisorbed configuration has a distance of ${\sim}$ 3.5 {\AA} distance between N and the Cu surface.

To properly learn the chemistry of adsorption, electrostatic interactions and charge transfer between the BTA molecule and metal surface must be considered. The chemisorption of the BTA molecule on Cu\hkl(111) surface involves charge transfer between N and Cu atoms. \autoref{fig:Cu-BTA}c and d compare the Hirshfeld charges on atoms predicted using DFT and NequIP-LR for the chemisorbed configuration while \autoref{fig:Cu-BTA}e and f compare the Hirshfeld charges on atoms for the  physisorbed configurations. The Cu atoms closest to the BTA molecule acquire a more negative charge compared to the rest. 

\autoref{fig:Cu-BTA}g and h visualize the force prediction errors for different atoms in the chemisorbed structure, comparing the performance of NequIP and NequIP-LR.
The results clearly show that NequIP-LR significantly reduces force errors, which is crucial for the correct evolution in MLIP-based molecular dynamics simulations. Notably, atoms near the BTA-Cu interface exhibit large force errors when using NequIP, whereas NequIP-LR achieves much lower errors. We attribute this improvement to the incorporation of charge distribution, which better differentiates interfacial atoms from the rest. This example structure highlights the importance of  MLIPs with charges for accurately predicting the properties of molecules chemisorbed or physisorbed on metal surfaces. Table \ref{table:rmse} shows that NequIP-LR achieves lower force and energy RMSE compared to NequIP.

\subsubsection*{Benzotriazole (BTA) on the Cu\hkl(111) surface with solvent}

Next, we investigate the BTA-Cu dataset in the presence of water as a solvent. The dataset contains two types of structures: BTAH molecules that exist under near neutral conditions and deprotonated form \ce{BTA-} that exists under high pH conditions (pH$>$8)\cite{gattinoni2015understanding,tanEnvironmentallySustainableCorrosion2022}. For reference, BTA has a pKa of 8.4 at 25$^{\circ}$C.
We trained NequIP and NequIP-LR on a dataset generated from ab initio MD simulations of BTAH and \ce{BTA-} on Cu\hkl(111) surface. The dataset consists of 300 simulation steps each for BTAH and \ce{BTA-}. In the dataset, BTAH is surrounded by water molecule, while \ce{BTA-} is accompanied by \ce{H3O+} ion as shown in Fig. S3a and b, respectively. As seen in Table \ref{table:rmse}, NequIP-LR outperforms NequIP in energy RMSE with 0.284 meV/atom compared to 0.709 meV/atom. In contrast, both models exhibit comparable accuracy in force RMSE, with values of 10.38 meV/{\AA} and 11.23 meV/{\AA}, respectively.

\subsubsection*{\ce{Ag3} cluster}

\begin{figure}[htbp]
    \centering
    \includegraphics[width=0.8\textwidth]{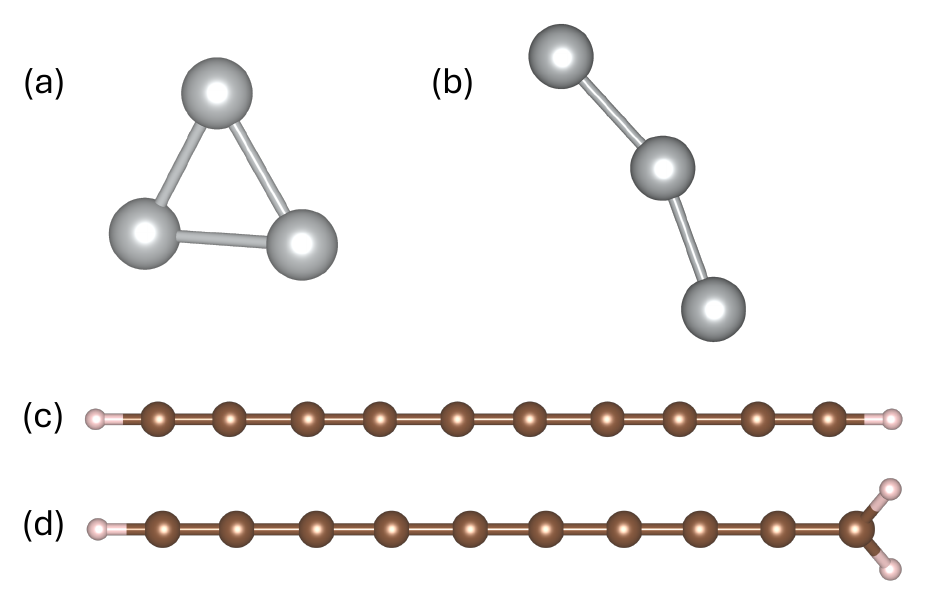}
    \caption{Non-periodic datasets used to benchmark the performance of NequIP-LR. \ce{Ag3} cluster in (a) +1 and (b) -1 total charge state. The ground state configuration is triangular in +1 charge state but trimer for -1 charge state. Structure of the carbon chain (a) C\(_\text{10}\)H\(_2\) and (b) C\(_\text{10}\)H\(_3^+\).}
    \label{fig:Ag3-cc}
\end{figure}

This dataset contains non-periodic \ce{Ag3} clusters in two different geometrical configurations~\cite{koFourthgenerationHighdimensionalNeural2021a}. The potential energy surface of small clusters is influenced by their ionization state, with the ground-state geometry varying depending on the total charge of the cluster~\cite{duanmu2016geometries,goel2012density,fournier2007trends,de2011effect}.
Unlike the previously discussed charge-neutral datasets, this dataset includes structures with different total charge states of +1 and -1. When the total charge is +1, the cluster adopts a triangular configuration, whereas for a charge of -1, it forms a trimer structure, as illustrated in \autoref{fig:Ag3-cc}a and b.

Due to the small system size, the entire structure lies within the typical cutoff radius used for MLIPs. Therefore, the challenge lies not in capturing long-range effects but in accurately modeling the distinct potential energy surfaces corresponding to different charge states.
Most MLIPs are not designed for this task since they lack information about total charge of the system.
On training NequIP and NequIP-LR on this dataset, we achieve lower force and energy RMSE using NequIP-LR compared to 4G-HDNNP by using a lower cutoff radius and less than 50\% of the data  (5000 structures) for training as seen in Table \ref{table:rmse}. This demonstrates that the combination of equivariance and charge equilibration considerably improves the model accuracy and data efficiency.

\subsubsection*{Carbon chain}

The carbon chain system shown in \autoref{fig:Ag3-cc}c and d exhibits long-range charge transfer induced by protonation, which changes the total charge and the local structure in parts of the system. 
The system consists of a chain of 10 \(sp\)-hybridized carbon atoms terminated by two hydrogen atoms (\ce{C10H2})
as shown in panel (c). Upon protonation of the terminal carbon atom, its hybridization state changes to \(sp^2\) and the electronic structure of the resulting C\(_\text{10}\)H\(_3^+\) cation shown in (d) is modified even at large distances from the proton.

We train our model on this dataset with less than 50\% of data (5000 out of total 10019 structures) compared to 4G-HDNNP. As demonstrated in Table \ref{table:rmse}, NequIP-LR achieves lower force and energy RMSE values compared to both 4G-HDNNP and NequIP. This highlights its superior ability to effectively capture long-range interactions. At first, it is surprising that NequIP achieves a lower force RMSE than 4G-HDNNP, despite being agnostic to the system’s charge. However, the two charge configurations differ in composition by a hydrogen atom, which allows NequIP to distinguish them. This explanation is supported by the fact that NequIP performs poorly on the \ce{Ag3} cluster dataset, where the compositions remain identical for +1 and -1 charge state. Nevertheless, this finding underscores the need for additional metrics to effectively compare the performance of MLIPs trained on different charge states.

The force and energy RMSEs for all datasets are summarized in Table \ref{table:rmse}. The model and training parameters are provided in the Supporting Information.

\begin{table}[htbp]
\centering
\begin{tabular}{|l|c|c|c|c|c|c|}
\hline
\multirow{2}{*}{Dataset} & \multicolumn{3}{c|}{Force RMSE (meV/{\AA})} & \multicolumn{3}{c|}{Energy RMSE (meV/atom)} \\ \cline{2-7}
                         & 4G-HDNNP & NequIP & NequIP-LR & 4G-HDNNP & NequIP & NequIP-LR \\ \hline
\ce{Au2}-MgO            & 66.00    & 34.32  & \textbf{21.99}  & 0.219    & 1.030  & \textbf{0.173}  \\ \hline
BTA-Cu                  & -        & 7.82   & \textbf{2.33}   & -        & 0.482  & \textbf{0.432}  \\ \hline
BTA(\ce{H2O})-Cu            & -        & \textbf{10.38}  & 11.23   & -        & 0.709  & \textbf{0.284}  \\ \hline
\ce{Ag3} cluster                & 31.69    & 2144.68*  & \textbf{21.01}*  & 1.323    & 498.394*  & \textbf{0.816}*  \\ \hline
Carbon chain            & 78.00    & 70.83*   & \textbf{54.59}*   & 1.194    & 1.333*   & \textbf{0.844}*   \\ \hline
\end{tabular}
\caption{Force RMSE (meV/{\AA}) and energy RMSE (meV/atom) for the datasets using 4G-HDNNP, NequIP, and the NequIP-LR model implemented in this work. * indicates that NequIP and NequIP-LR were trained on less than 50\% of the data used for 4G-HDNNP.}
\label{table:rmse}
\end{table}

\subsection*{Hyperparameters, Data Efficiency, and Computational Complexity}

\begin{figure}[htbp]
    \centering
    \includegraphics[width=0.8\textwidth]{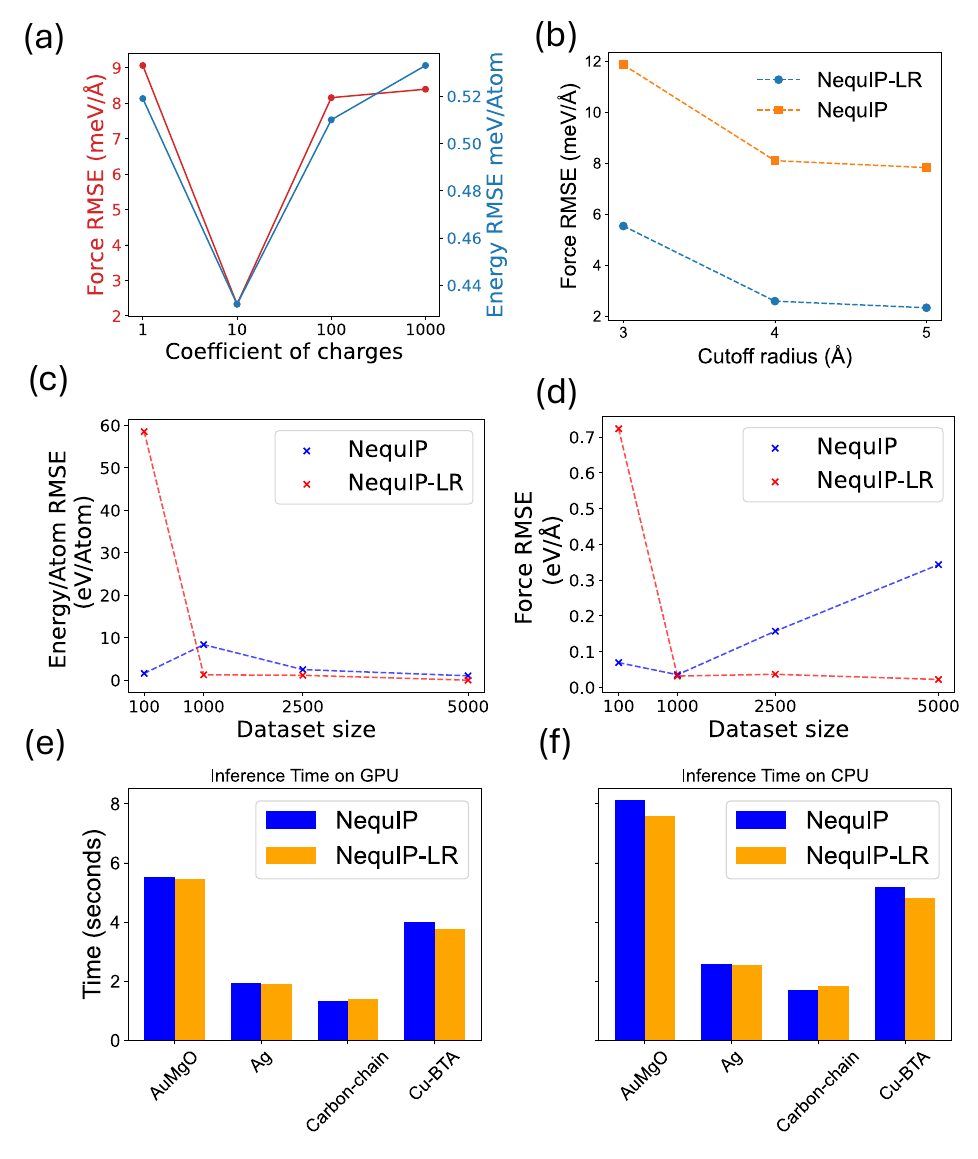}
    \caption{(a) Variation of force and energy RMSE with the coefficient of charge error in the loss function, $\alpha_Q$ for the BTA-Cu dataset.  (b) Force RMSE as a function of cutoff radius with a fixed number of interaction layers (6) for NequIP and NequIP-LR for the BTA-Cu dataset. NequIP-LR achieves superior performance with a cutoff radius of just 3 {\AA}, compared to  NequIP with a 5 {\AA} cutoff. Learning curve for (c) energy RMSE and (d) force RMSE for NequIP and NequIP-LR on \ce{Au2} cluster on MgO dataset. 
    Inference time for (e) GPU and (f) CPU using NequIP and NequIP-LR MLIPs. }
    \label{fig:general}
\end{figure}

An important hyperparameter that determines the accuracy of NequIP-LR and MLIPs with charges is the coefficient of charge error in the loss function, $\alpha_Q$. Hence, we examined its impact on the model accuracy for BTA-Cu dataset. An increase in the coefficient value places greater emphasis on reproducing the atomic charges correctly. \autoref{fig:general}a shows the variation in force and energy RMSE as a function of coefficient of charges error in the loss function. We found that $\alpha_Q=10$ gives a fair tradeoff between the errors in forces, energies, and charges (shown in Fig. S1), hence, this value was used for reporting all results in this work. 

To examine the relationship between model parameters and accuracy, we conducted hyperparameter screening tests on the cutoff radius used in the model. \autoref{fig:general}b shows the variation in force RMSE as a function of cutoff radius on BTA-Cu dataset. Our results indicate that NequIP-LR achieves higher accuracy with a smaller cutoff radius of 3 {\AA} compared to NequIP with a 5 {\AA} cutoff. Specifically, NequIP gives a force RMSE of 7.82 meV/{{\AA}} and an energy RMSE of 0.482 meV/atom and NequIP-LR reduces these errors to 5.53 meV/{\AA} and 0.241 meV/atom, respectively, even with a smaller 3 {\AA} cutoff. This improvement makes NequIP-LR well-suited for large-scale MD simulations, as it provides more accurate force predictions at a reduced computational cost due to lower number of message-parsing operations.

To assess data efficiency, we plot the learning curves for the \ce{Au2} cluster on the MgO dataset, comparing NequIP and NequIP-LR using energy and force RMSE, as shown in \autoref{fig:general}c and d, respectively. The RMSE for NequIP-LR decreases monotonically, whereas NequIP does not exhibit a similar trend. This discrepancy may stem from NequIP’s higher sensitivity to specific structures within the dataset. Additionally, when the dataset size is small (100 samples), NequIP-LR exhibits a higher error than NequIP, likely due to its greater model complexity, which makes it more prone to overfitting in low-data regimes. Hence, a dataset size of 1000 is required for  achieving a reasonable accuracy with NequIP-LR.

While we have demonstrated NequIP-LR’s accuracy across diverse datasets, its more complex architecture results in longer training times compared to NequIP, primarily because it requires more epochs to converge on the same dataset. Once trained, the key metric related to computational complexity for practical applications such as MD simulations is the inference time per energy and force evaluation.  \autoref{fig:general}e and f compare the average inference time for energy and force evaluations using 20 structures from each of the datasets. Despite incorporating additional features such as long-range interactions and charge equilibration, NequIP-LR requires comparable or lower inference time than NequIP. This is because we ensure that both models have the same number of convolutional layers despite having two sets of equivariant neural networks in NequIP-LR.
This favorable computational complexity, combined with its superior accuracy, makes NequIP-LR well-suited for large-scale MD simulations.

\section*{Discussion}
We have developed a new MLIP that equips equivariant models with additional capabilities: long-range interactions and global charge distribution among atoms.  While we implemented and tested our framework using NequIP, it can be similarly extended to other equivariant models. 
Our framework offers three key advantages. First, it enhances the accuracy of force and energy predictions in heterogeneous systems by effectively distinguishing atoms based on the   charge distribution.  Unlike methods that rely solely on local environments, our framework systematically determines atomic charges through a global charge equilibration process, ensuring a more physically consistent representation of charge distribution. Hirshfeld charges serve as a good proxy for the charge distribution and enhance the accuracy of the force and energy predictions.
Second, it incorporates long-range interactions and charge transfer beyond the cutoff radius, which are essential for accurately modeling many complex systems. With short-range MLIPs, modeling long-range interactions may require a large number of convolutional layers or a higher cutoff radius, both of which increase the computational cost significantly. Lastly, our MLIP can model the potential energy surface of a given structure across different total charge states, making it useful for studying, for example, charged defects in materials.

We anticipate that this MLIP will be particularly useful for large-scale MD simulations of heterogeneous systems where long-range interactions beyond the cutoff radius and charge transfer play a significant role. Examples include interfaces with charge transfer encountered in areas such as catalysis and electrochemistry. 
By accurately simulating nonlocal charge transfer across the system through charge equilibration, our approach offers the potential for investigating and uncovering chemical and physical phenomena including reactions and transport occurring at interfaces. We  envision that further advancements in MLIPs will be enabled by explicit treatment of charge distributions, similar to our developed MLIP, bringing them closer to quantum mechanical methods.

\section*{Methods}
 \subsection*{Charge equilibration scheme}

Similar to 4G-HDNNP, we represent the atomic charges by Gaussian charge densities of width \(\sigma_i\) equal to the covalent radii of the chemical elements. Here, we show the charge equilibration \cite{rappeChargeEquilibrationMolecular1991} procedure used to determine the charge distribution.

In this scheme, the energy expression is minimized to distribute charges among the atoms.
\begin{equation}
    E_{\text{Qeq}} = E_{\text{lr}} + \sum_{i=1}^{N_{\text{at}}} \left( \chi_i Q_i + \frac{1}{2} J_i Q_i^2 \right) 
\end{equation}
 
Here, \(E_\text{lr}\) is the long-range electrostatic energy of the Gaussian charges $Q_i$ while $\chi_i$ and $J_i$ are the electronegativity and hardness of atom $i$, respectively. $N_{\text{at}}$ is the total number of atoms in the structure. For a system with Gaussian charges, the long-range electrostatic energy can be obtained analytically as

\begin{equation}
    E_{\text{lr}} = 
\sum_{i=1}^{N_{\text{at}}} \sum_{j}^{N_{\text{at}}} 
\frac{\text{erf}\left(\frac{r_{ij}}{\sqrt{2}\gamma_{ij}}\right)}{r_{ij}} Q_i Q_j 
+ \sum_{i=1}^{N_{\text{at}}} \frac{Q_i^2}{2 \sigma_i \sqrt{\pi}}
\end{equation}

where 
\begin{equation}
    \gamma_{ij} = \sqrt{\sigma_i^2 + \sigma_j^2}
\end{equation}

The atomic charges are obtained by minimizing the energy with respect to the charges, giving a set of linear equations for $Q_i$

\begin{equation}
    \frac{\partial E_{Qeq}}{\partial Q_i} = 0, \forall i = 1, ..., N_{at} \implies \sum_{j=1}^{N_{at}} A_{ij} Q_j + \chi_i = 0
\end{equation}

where the matrix \(\textbf{A}\)  is given by, 

\begin{equation}
     {[\mathbf{A}]_{ij}}  = \begin{cases}
    J_i + \frac{1}{\sigma_i \sqrt{\pi}}, & \text{if } i = j \\
    \frac{\text{erf}\left(\frac{r_{ij}}{\sqrt{2} y_{ij}}\right)}{r_{ij}}, & \text{otherwise}
\end{cases}.
\end{equation}

In addition, the charges  $Q_i$  must satisfy the constraint that the summation of all charges equals the total charge  $Q_\text{tot}$. This constraint can be incorporated into the solution for  $Q_i$  by formulating it as a linear system with a Lagrange multiplier,  $\lambda$ 

\begin{eqnarray}
    \begin{pmatrix}
{\mathbf{A}} & \vline & \begin{matrix} 1 \\ \vdots \\ 1 \end{matrix} \\
\hline
\begin{matrix} 1 & \cdots & 1 \end{matrix} & \vline & 0
\end{pmatrix}
\begin{pmatrix}
Q_1 \\ \vdots \\ Q_{N_{\text{at}}} \\ \hline \lambda
\end{pmatrix}
=
\begin{pmatrix}
-\chi_1 \\ \vdots \\ -\chi_{N_{\text{at}}} \\ \hline Q_{\text{tot}}
\end{pmatrix}
\end{eqnarray}

The linear system is solved to obtain the atomic partial charges $Q_i$. 

\noindent The code for reproducing the simulations in this manuscript is available at \url{https://github.com/ahmad-research-group/nequip-charge/tree/charge-encoding}
 
\begin{acknowledgement}
We thank Jörg Behler for helpful discussions.  This work was supported by Samsung Advanced Institute of Technology Global Research Outreach program.
We acknowledge Lonestar6 research allocations (DMR23017 and DMR24003) at the Texas Advanced Computing Center (TACC) for providing computational resources that have contributed to the research results reported within this paper.

\end{acknowledgement}

\begin{suppinfo}

Details of training and testing procedure, parity plot of forces and energies.

\end{suppinfo}

\bibliography{zotero,manual-refs}



\section*{Supplementary material}

# Supporting Information:

# Equivariant Machine Learning Interatomic Potentials with Global Charge Redistribution

Moin Uddin Maruf,[†] Sungmin Kim,[‡] and Zeeshan Ahmad[*,†]

[†]Department of Mechanical Engineering, Texas Tech University, Lubbock, Texas 79409, USA

[‡]Samsung Advanced Institute of Technology, Samsung Electronics, Suwon 16678, Republic of Korea

E-mail: zeeahmad@ttu.edu

# 1 Supplementary Figures

Both the BTA on Cu(111) surface datasets were generated in this work. $Au_2$ cluster on MgO(001) surface, $Ag_3$ metal clusters and Carbon chain dataset are from Ko et al.[1].

### 1.0.1 Benzotriazole (BTA) on Cu(111) surface

Table S1

| - | NequIP | NequIP-LR |
|---|---|---|
| No. of epochs | 85 | 2000 |
| Batch size | 1 | 1 |
| Cutoff radius | 5 | 5 |
| number of layers | 6 | 6 |
| $l_{max}$ | 1 | 1 |

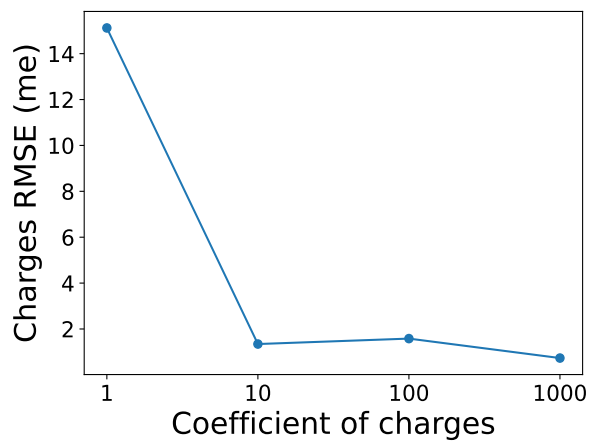

Figure S1: Charges RMSE with the coefficient of charge error used in the loss function ($\alpha_q$).

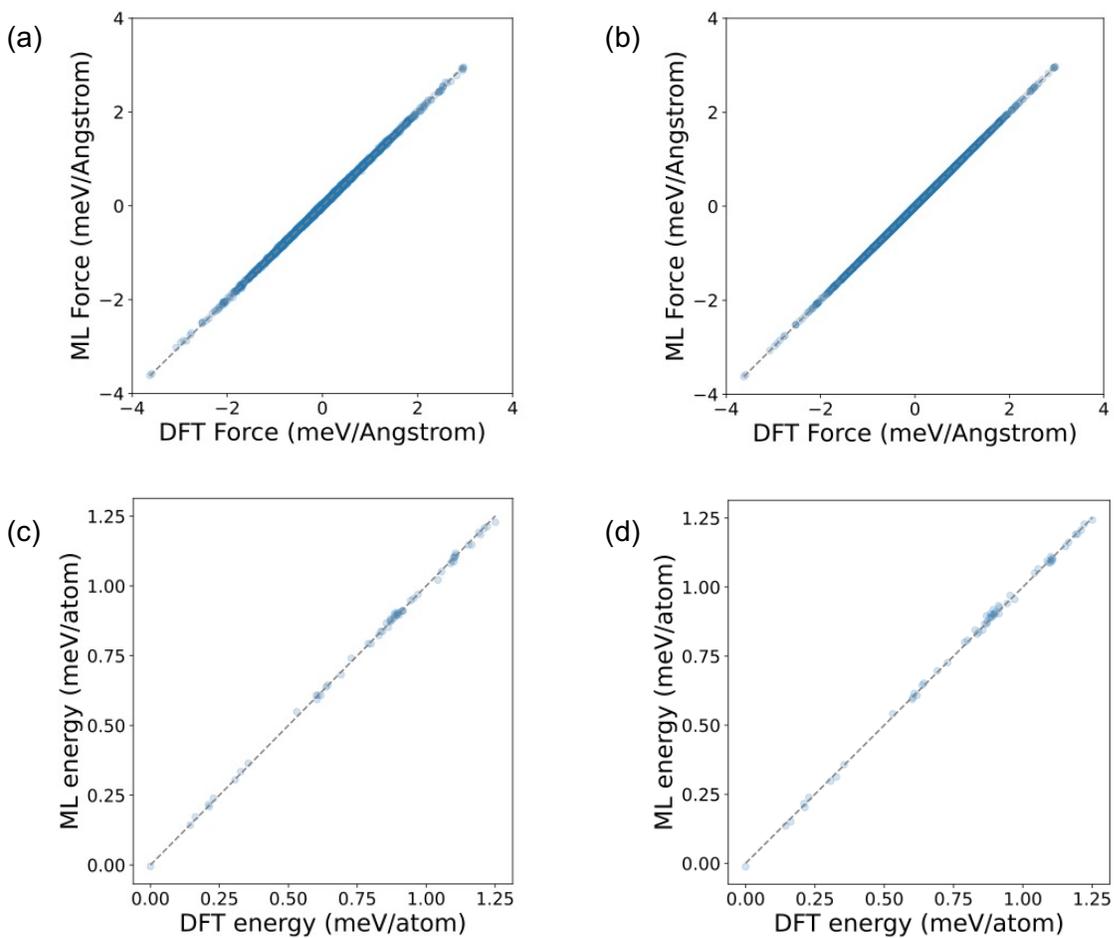

Figure S2: Parity plot of forces obtained from (a) NequIP, and (b) NequIP-LR, energies from (c) NequIP, and (d) NequIP-LR models on Benzotriazole (BTA) on Cu(111) surface dataset.

### 1.0.2 Benzotriazole (BTA) on the Cu(1 1 1) surface with solvent

Table S2

| -                | NequIP | NequIP-LR |
|------------------|--------|-----------|
| No. of epochs    | 112    | 171       |
| Batch size       | 1      | 1         |
| Cutoff radius    | 4      | 4         |
| number of layers | 6      | 6         |
| $l_{\max}$       | 1      | 1         |

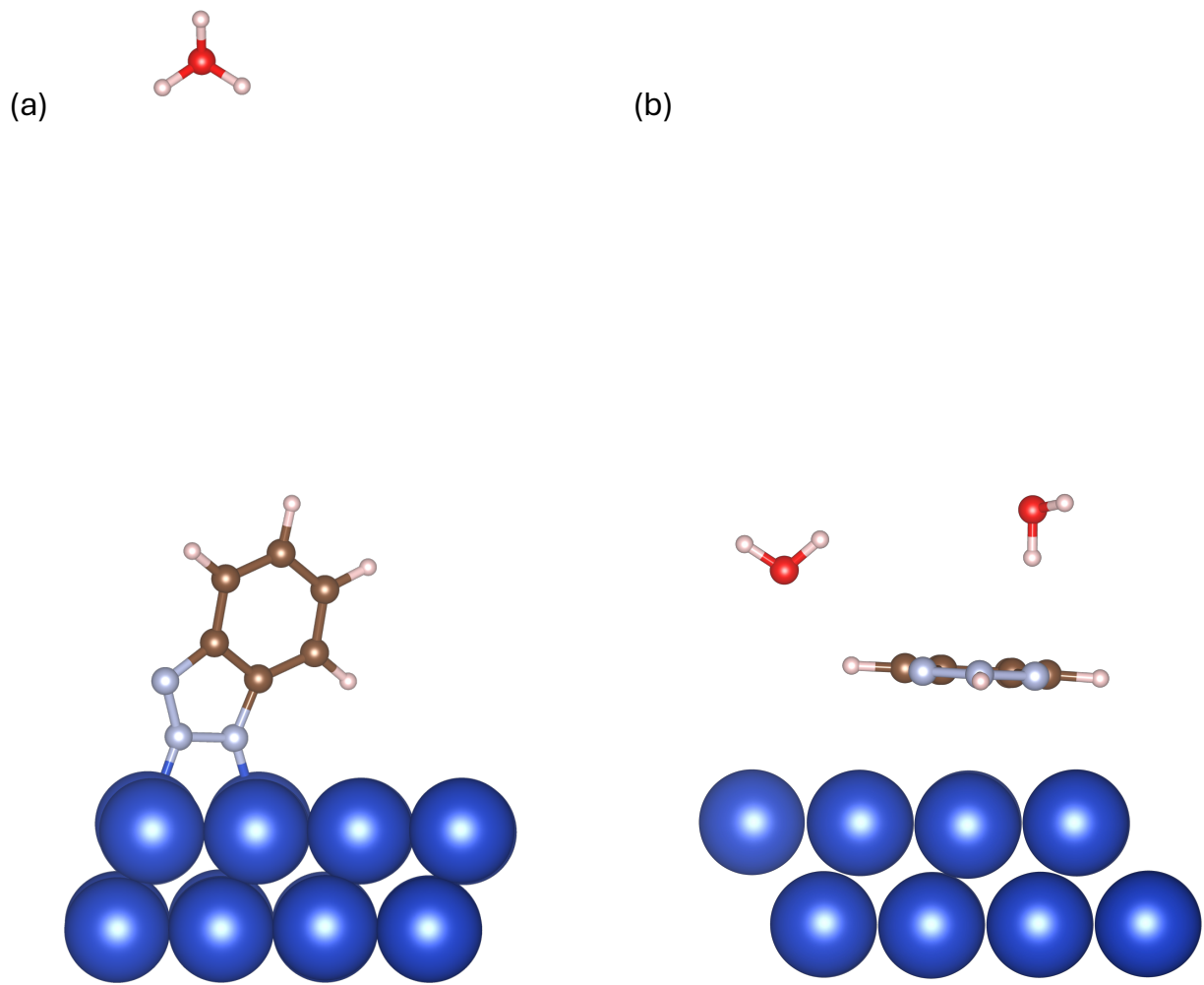

Figure S3: Structures of (a) BTA$^-$, the deprotonated state that exists under high pH and (b) BTAH that exists under near neutral conditions on Cu (1 1 1) surface.

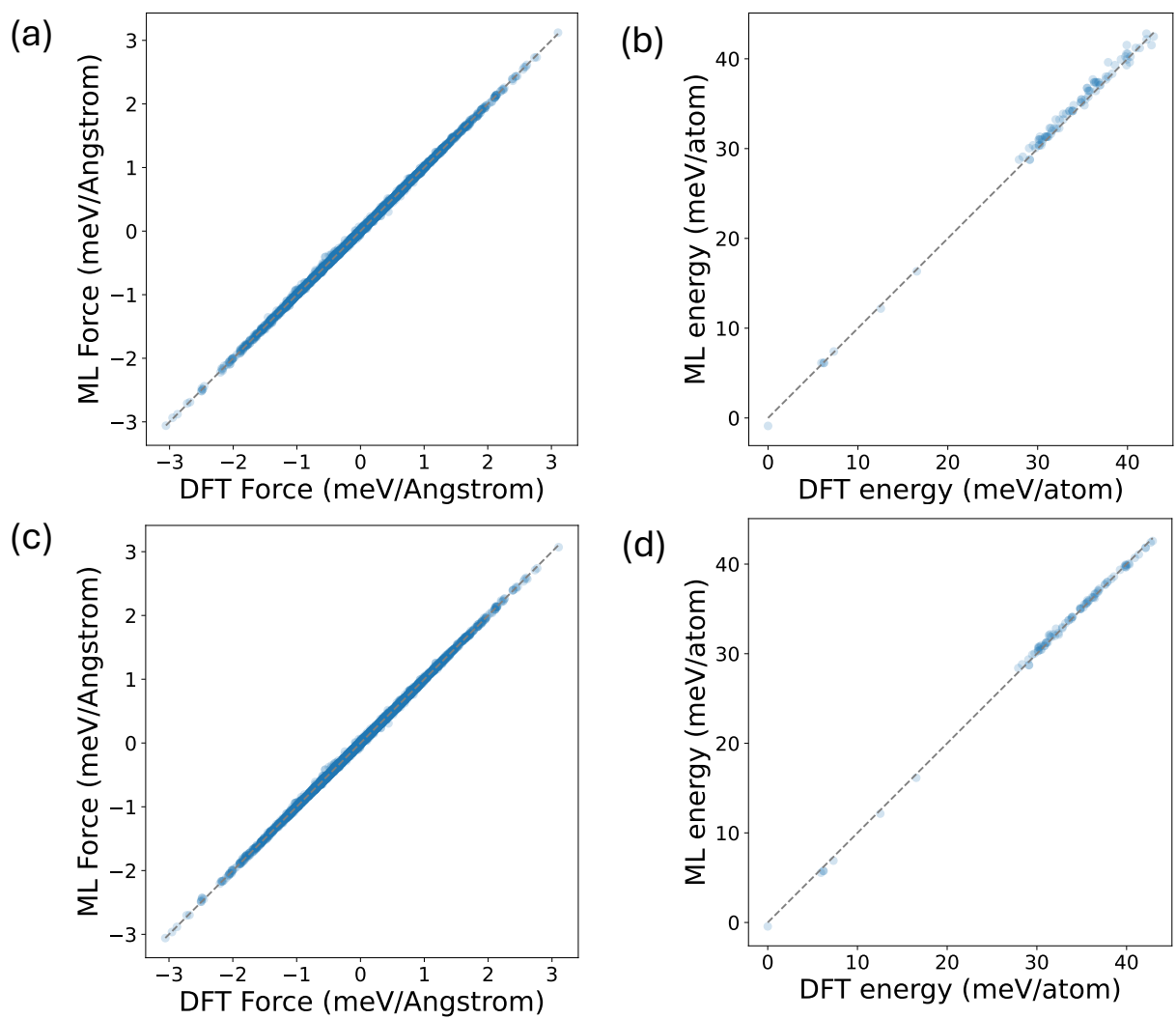

Figure S4: Parity plot of forces obtained from (a) NequIP, and (c) NequIP-LR, energies from (b) NequIP, and (d) NequIP-LR models on Benzotriazole (BTA) on the Cu(1 1 1) surface with solvent.

### 1.0.3 Au$_2$ cluster on MgO(001)

Table S3

| -                | NequIP | NequIP-LR |
|------------------|--------|-----------|
| No. of epochs    | 156    | 1056      |
| Batch size       | 15     | 10        |
| Cutoff radius    | 5.5    | 5.5       |
| number of layers | 6      | 6         |
| $l_{\max}$       | 2      | 2         |

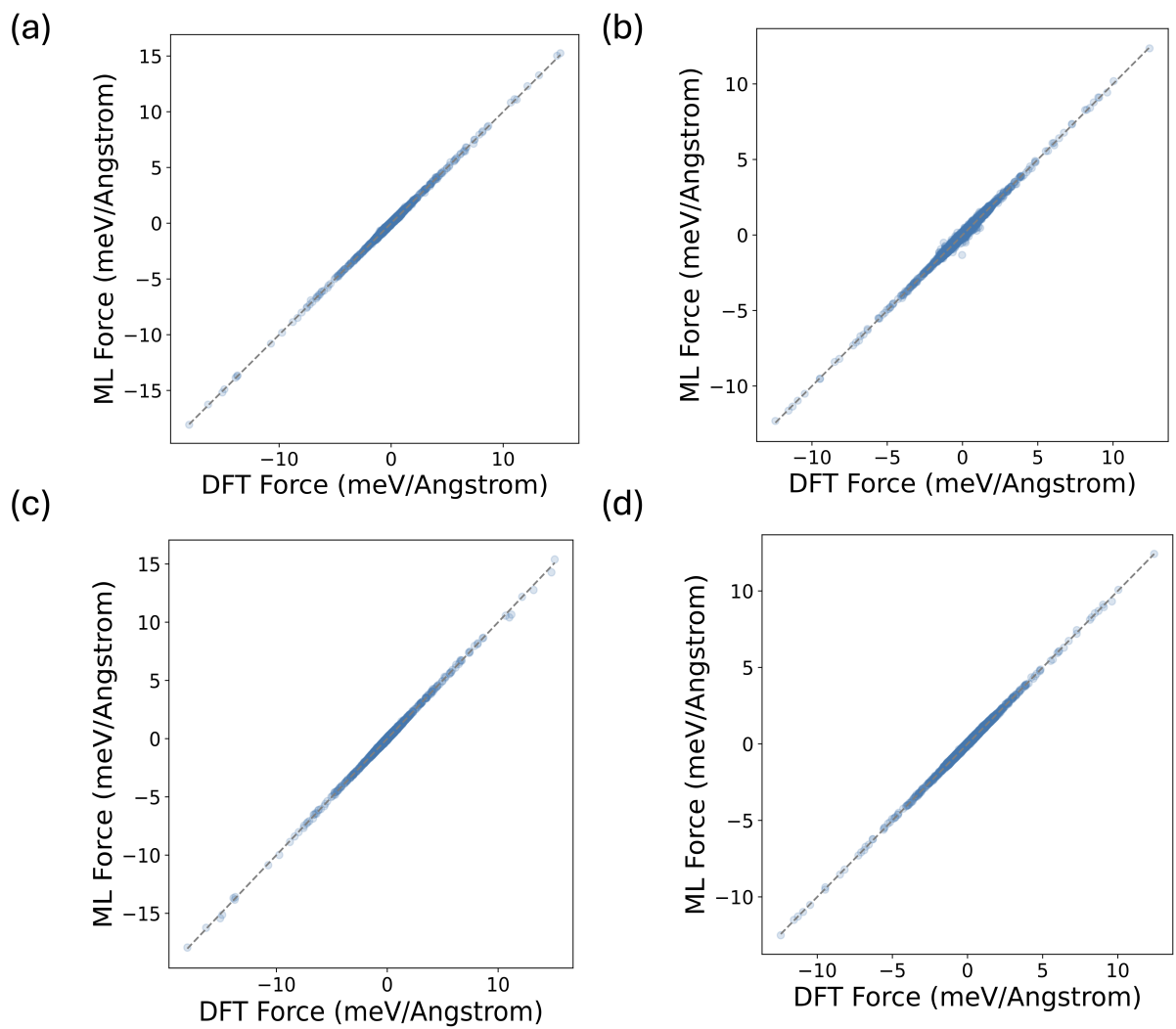

Figure S5: Parity plot of forces obtained from (a and b) NequIP and (c and d) NequIP-LR models for undoped and doped structures on $Au_2$ cluster on MgO(001) dataset, respectively.

### 1.0.4 Metal clusters: Ag$_3$

Table S4

| - | NequIP | NequIP-LR |
|---|---|---|
| No. of epochs | 280 | 1596 |
| Batch size | 15 | 100 |
| Cutoff radius | 5.29 | 5.29 |
| number of layers | 8 | 8 |
| $l_{\max}$ | 2 | 2 |

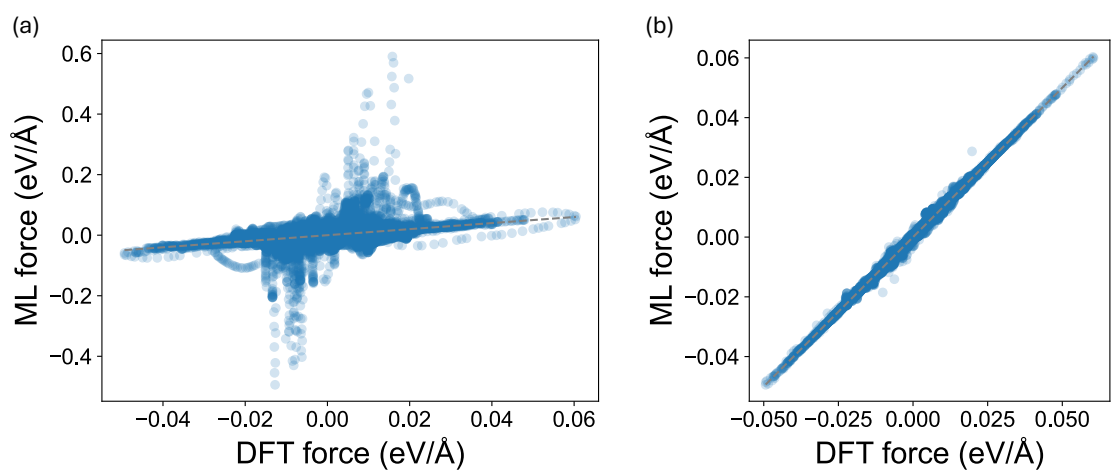

Figure S6: Parity plot of forces obtained from (a) NequIP and (b) NequIP-LR models on Ag$_3$ metal clusters dataset.

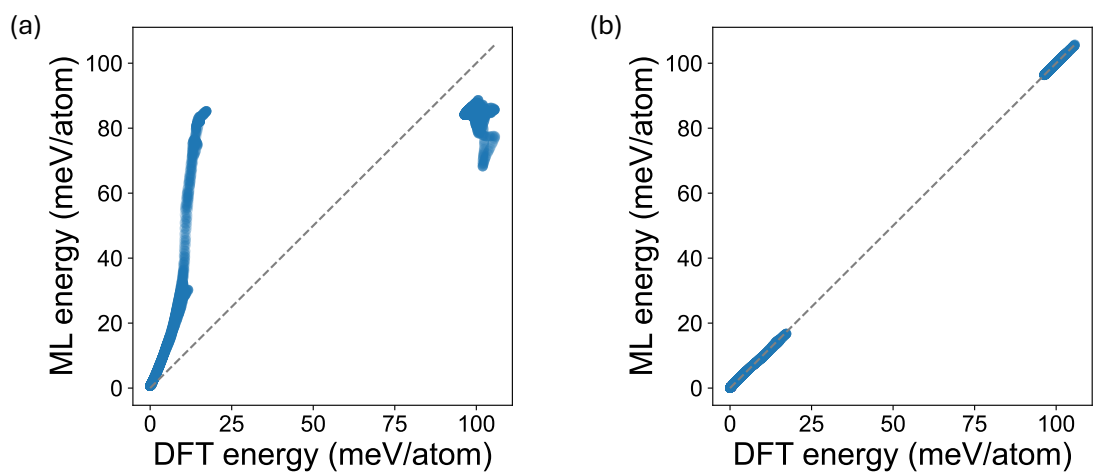

Figure S7: Parity plot of energies obtained from (a) NequIP and (b) NequIP-LR models on $Ag_3$ metal clusters dataset.

### 1.0.5 Carbon Chain

Table S5

| -               | NequIP | NequIP-LR |
|-----------------|--------|-----------|
| No. of epochs   | 681    | 2517      |
| Batch size      | 50     | 100       |
| Cutoff radius   | 5      | 5         |
| number of layers| 8      | 8         |
| $l_{\max}$      | 1      | 1         |

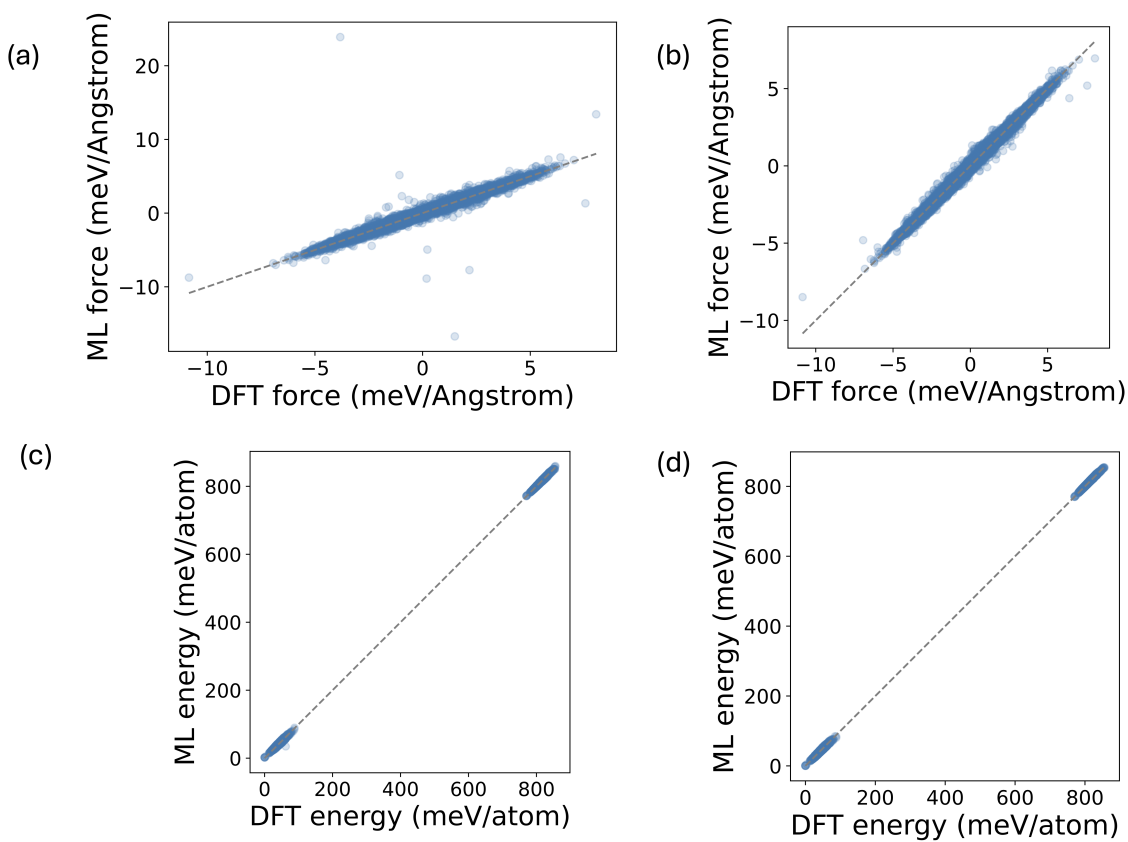

Figure S8: Parity plot of forces obtained from (a) NequIP and (b) NequIP-LR, energies from (c) NequIP and (c) NequIP-LR models on Carbon chain dataset.


\end{document}